\documentclass[english,prl,twocolumn]{revtex4}
\usepackage[toc,page,title,titletoc,header]{appendix}
\usepackage[T1]{fontenc}
\usepackage[latin1]{inputenc}
\usepackage{graphicx}
\usepackage{amsmath}
\usepackage{shapepar}

\usepackage{color}

\usepackage{amssymb, amsbsy, amsthm}
\makeatletter

\usepackage{babel}
\makeatother
\begin{document}
\title{Uncertainty-Complementarity Balance  as a General Constraint on Non-locality}
\author{Liang-Liang Sun,   Sixia Yu\footnote[1]{email: yusixia@ustc.edu.cn}, Zeng-Bing Chen\footnote[2]{email: zbchen@ustc.edu.cn}
}
\affiliation{Hefei National Laboratory for Physical Sciences at Microscale and Department
of Modern Physics, University of Science and Technology of China, Hefei,
Anhui 230026}
\date{\today{}}
\begin{abstract}
    We propose an  uncertainty-complementarity balance relation and build quantitative connections among  non-locality, complementarity, and uncertainty. Our balance relation, which is formulated in a theory-independent manner, states that for two measurements performed sequentially, the complementarity demonstrated in the first measurement in terms of disturbance is no greater than the uncertainty of the first measurement. Quantum theory respects our balance relation, from which the Tsirelson  bound can be derived, up to an inessential assumption.  In the simplest Bell scenario, we show  that  the  bound of Clauser-Horne-Shimony-Holt inequality for a general non-local theory can be expressed as a function of the balance strength, a constance for the given theory.  As an application, we derive the balance strength as well as the nonlocal bound of Popescu-Rohrlich box.   Our results shed light on quantitative connections among three fundamental concepts, i.e., uncertainty, complementarity and non-locality.
 \end{abstract}


\maketitle

{\it Introduction--- }The core formulation of quantum mechanics (QM) is based on the structures of Hilbert space, which gives rise to fundamental non-classical features such as uncertainty, complementarity, and non-locality. However, it is still an open question with respect to the underlying physical principles behind the structures of Hilbert space. This is quite unlike other successful theories such as the theory of relativity and statistical mechanics,  whose formalisms  can be directly derived  from several fundamental physical principles.  One fruitful approach to tackle the problem is to trace various quantum features  back to its physical principles~\cite{hall,sp,t1,t2,gpt,nb,gs,Barr,tn1,tn2,tn3} in a theory-independent manner.
For examples, Bell has provided a general framework to quantify non-locality~\cite{bel}, within which the observed bipartite correlations  precludes local realistic models of QM.
Barrett introduced a framework  applicable  to generalized probabilistic theories~\cite{t1,t2,gpt,nb,gs,Barr} in which, some  properties thought  special to QM, $e.g.$, teleportation ~\cite{ta,te},  purification~\cite{pur}, coherence\cite{woc},  and entanglement swapping~\cite{sw},  have been found to be not confined to QM.

Actually, most of our understandings of QM in an axiomatic manner are gained by singling out QM based on the  non-locality demonstrated by the correlations of compatible measurements. Many efforts have been devoted to understanding why quantum correlation is strong enough  to be nonlocal but not so strong to be maximum  nonlocal~\cite{ns}. For example,  the Clauser-Horne-Shimony-Holt (CHSH) inequality has an upper bound 2 for any local realistic theory,  the Tsirelson bound  $2\sqrt{2}$ for QM~\cite{bel}, and reaches its maximum   4 for  the Popescu-Rohrlich box model (PR-box)~\cite{pr}.   Various principles have been proposed, $e.g.$, the information causality principle~\cite{ic},    nontrivial communication complexity~\cite{wvd},  global exclusion principle~\cite{cl,by,ca}, to explain the quantum mechanical non-local bound.  While these results have gained some valuable insights  to this question, they do not explain in a quantitative manner the intrinsic complementarity and uncertainty that are present in any general nonlocal theory~\cite{gpt,nb,gs,Barr,ns,hor}. So far, the complementarity is taken only intuitively  as a necessary condition for a non-local theory  to respect the no-instantaneous communication principle~\cite{tn1}. Hence,  a natural question arises as to  whether one can quantitatively determine the  nonlocality bound with  uncertainty and complementarity for a general theory, and then explain the Tsirelson bound.

In this Letter we give an answer to the above question by introducing an uncertainty-complementarity balance relation which is shown to impose  a strong constraint on  non-locality. For this purpose we consider the scenario in which two measurements, which might not be compatible to each other in comparison with compatible measurements in the usual Bell scenario, are performed sequentially. Our balance relation states that  the complementarity  demonstrated in the second measurement in terms of disturbance is no greater than the  uncertainty presented in the first measurement. The balance relation is shown to be respected by QM and can account for the Tsirelson bound, together with an additional assumption. Essential in our balance relation there is a constance called balance strength for each specific theory. It turns out that the balance constance is intimately related to the  non-locality: the  bound  of $\operatorname{CHSH}$ inequality  for a specific theory can be casted into  a function of  its balance  strength, from which  we reproduce the non-local bound of PR-box as an application.  Vise versa,  non-locality displayed by a theory also imposes a constraint on the possible values of its balance strength.

\emph{Uncertainty and Complementarity balance --- } To start with, we assume that a given physical system can be in different states, which are nothing else than some mathematical structures that help to determine the statistics of all possible measurements performed on the system. For a given state of the system, each measurement, e.g., $A$, will result in a probability distribution, e.g., $\{p(a|A)\}$, over all possible outcomes of the measurement. Uncertainty of the measurement $A$ can be quantified in various ways and here we shall take
\begin{equation}
\delta_{A}=\left(\textstyle\sum_{a}\sqrt{p({a|A})}\right)^{2}-1
\end{equation}
as the measure of uncertainty.  Specially, if observable $A$ is a two-outcome observable with assigned values $\{0,1\}$,  we have
$\delta_{A}^2=4p(0|A)p(1|A)=1-\bar {A}^{2}$ with $\bar A=p(0|A)-p(1|A)$ being the expectation of the observable $A$.

Complementarity states that there are pairs of incompatible properties that cannot be measured simultaneously. Consequently one measurement might cause inevitable disturbance to a later incompatible measurement. Thus we consider two incompatible measurements performed in sequel. Here we consider only sharp measurements, measurements that are accurate and repeatable~\cite{sm1,sm2}, for the sake of clarity. After a sharp measurement $A$, depending on the outcome $\mu$, the system might be brought into some states~$\mathcal{S}_{\mu|A}$ that might be different from the original stats with probability $p({\mu|A})$. This state is conveniently referred to as the ``eigenstate'' of observable $A$. On average the output state after the measurement $A$ would be an ensemble $\{ p({\mu|A}), \mathcal{S}_{\mu|A}\}$ of eignestates. And then the second measurement $A^\prime$ is performed on this ensemble, resulting in a probability distribution
 \begin{equation}\label{did}
p_{a|A\to A'}=\sum_{\mu}p({\mu|A})P({a|A',\mathcal{S}_{\mu|A}}),
\end{equation}
where   $P({i|A',\mathcal{S}_{\mu|A}})$ denotes  the probability of obtaining $i$ when measuring $A'$ on $\mathcal{S}_{\mu|A}$. Had the first measurement of $A$ not been performed the measurement of $A^\prime$ on the original state would give rise to the statistics $\{p({a^\prime|A^\prime})\}$ instead. Therefore
the difference between these two probability distributions
 \begin{eqnarray}\label{dis}
D_{A\rightarrow A'}=\sum _{a}\big|p({a|A'})-p({a|A\to A'})\big| \label{seq}
\end{eqnarray}
quantifies naturally the disturbance of the second measurement ($A^\prime$) caused by the first measurement ($A$) and thus provides a quantitative measure of complementarity.
We note that the  above measures of uncertainty  and disturbance  involve only probabilities thus they do not rely on specific structure of potential theory.

Quantum mechanically, the relations between uncertainty and complementarity have been extensively studied in terms of   measurement-disturbance relations ~\cite{qu1,qu2}, where the complementarity is commonly lower-bounded by a function of uncertainty~\cite{sss}.  On the other hand, since  a measurement with zero uncertainty can not  lead to a non-zero disturbance to a subsequent measurement, the complementarity should also impose some kinds of constraints on possible uncertainty of the first measurement. Quantum mechanically,  a projective  measurement $A^\prime$ is preformed after another projective  measurement $A$ it holds (See SM)
 \begin{eqnarray}\label{qmr}
\delta_{A}\geq D_{A\rightarrow A'},
\end{eqnarray}
meaning that  a measurement $A$ would not lead a  disturbance $D_{A\rightarrow A'}$ (to a following measurement $A'$) greater than its uncertainty $\delta_{A}$. We note that the quantum mechanical balance relation Eq.(\ref{qmr}) can be saturated:  $\delta_{A}=1$ and  $D_{A\rightarrow A}=1$ for a qubit in the  state  $\rho=|0\rangle\langle 0|$ with two measurements $A=\sigma_x$ and  $A'=\sigma_z$, then .

Now we are in the position to introduce a general balance relation similar to Eq.(\ref{qmr}) for any probabilistic theory. By denoting
\begin{equation}
\alpha=\sup\{\alpha^\prime\ge0|\delta_{A}\geq \alpha'D_{A\rightarrow A'}\ (\forall A, A', \mathcal S)\}
\end{equation} our generalized  balance relation  reads
\begin{eqnarray}\label{aaa1}
\delta_{A}\geq \alpha D_{A\rightarrow A'}.
\end{eqnarray}
The balance strength  can be taken as a benchmark parameter which reflects the relation between uncertainty and complementarity   just like  the maximum violation of Bell's inequalities for non-locality.   As shown above,  the balance strength for QM is 1, i.e., $\alpha_{\rm qm}=1$, since the balance relation Eq.(\ref{qmr}) is respected by QM and can be saturated. Roughly speaking, the balance strength quantifies the maximal violation to the quantum mechanical balance relation Eq(\ref{qmr}).

Our balance relation deals with a different scenario   from the Bell scenario that is considered  in the existing  principles such as  information causality principle,    nontrivial communication complexity,  and global exclusion principle.
In the Bell scenario one considers only the correlations of compatible, e.g., space-like separated, measurements and thus a kind of correlation strength is characterized.  In our balance relation we consider also incompatible measurements in a causal order,  giving rise to a balance  strength.

\emph{Non-locality under Uncertainty and Complementarity balance  --- }  We shall now show that the balance relation Eq.(\ref{aaa1}) will give rise to a constraint on non-locality.  Non-locality is commonly  quantified by the violations to some Bell inequalities such as the CHSH inequality~\cite{bel}. In this simplest Bell scenario two space-like separated observers Alice and Bob perform locally some two-outcome measurements.  Alice  can randomly measure one of two observables $A_0$ or $A_1$, and Bob can measure observables $B_0$ or $B_1$  with two outcomes labeled by $ \{0,1\}$. Denoting by $p(a,b|A_\mu,B_\nu)$ the probability of obtaining outcome $a,b\in\{0,1\}$ when Alice measure observable $A_\mu$ and Bob measures observable $B_\nu$ with $\mu,\nu=0,1$  the following CHSH inequality  \cite{bel} holds for any local realistic theory
\begin{eqnarray}
\mbox{CHSH}:=\sum_{a,b,\mu,\nu=0}^1(-1)^{a+b+\mu\nu}p(a,b|A_\mu, B_\nu)\leq2.
\end{eqnarray}
  To proceed we need to introduce a physical constraint on the transition probabilities $P(i|A^\prime,\mathcal S_{\mu|A})$ (appearing in Eq.(\ref{did})) obtained by measuring $A^\prime$ after the measurement of $A$ has been performed with outcome $\mu$, with $A,A^\prime$ being two arbitrary sharp measurements which are incompatible. The constraint reads
   \begin{eqnarray}\label{1}
\sum_{\mu}P(i|A',\mathcal{S}_{\mu|A})=1\quad (\forall i)
\end{eqnarray}
which assumes that the measurement of observable $A'$ on the ensemble $  \{\frac{1}{d}, \mathcal{S}_{\mu|A}\}$ (where $d$ denotes the number of possible outcomes of $A$)  yields an unbiased probability distribution. In QM this unbias assumption is satisfied and it is weaker than the symmetry relation $P(i|A',\mathcal{S}_{\mu|A})=P(\mu|A',\mathcal{S}_{i|A})=\operatorname{Tr}(|i\rangle_{A'}\langle i|\mu\rangle_{A}\langle\mu|)$.

 In the case of two measurements $A_0$ and $A_1$ for Alice we obtain (See SM)  the following constraint on the maximal violation to the CHSH inequality, i.e., nonlocality upper bound,
 \begin{equation}
\operatorname{CHSH}\leq \max_{\gamma,\tau} n_{\gamma,\tau} \label{fia}
\end{equation}
with
$
n_{\gamma\tau}=2\sqrt{f(\alpha,\gamma, \tau)}+2\sqrt { {f(\alpha,-\gamma, -\tau)}}
$
from the balance relation Eq.(\ref{aaa1}),
where\begin{widetext}
\begin{equation}\label{f}
f(\alpha,\gamma, \tau)=\frac{\alpha^2(\tau^{2}+\gamma^{2}-2)+2}{\big(\alpha^2(1+\gamma)^{2}+1\big)\big(\alpha^2(\tau^{2}-1)+1\big)
     +\big(\alpha^2(1+\tau)^{2}+1\big)\big(\alpha^2(\gamma^{2}-1)+1\big)},
     \end{equation}
\end{widetext}
in which
\begin{equation}\label{gt}
\gamma=1-2P(1|A_1,\mathcal{S}_{1|A_0}),\quad \tau=1-2P({1|A_0,\mathcal{S}_{1|A_1}})
\end{equation}
Since in the case of $\alpha>1$ the balance relation Eq.(\ref{qmr}) for QM is not violated, we consider in the following the case $\alpha\le 1$. For a given $\alpha$ a typical diagram of the upper bound of CHSH as a function of $\gamma$ and $\tau$ is plotted in Fig.1, showing that the maximal values are attained in the case of $\gamma=\tau$. Thus the upper bound Eq.(\ref{fia}) becomes
\begin{equation}
\mbox{CHSH}\le \max_\gamma n_\gamma,\quad n_\gamma=n_{\gamma\gamma}.
\end{equation}
The  the upper bound $\max n_\gamma$ is plotted in Fig1 as a function of the balance strength $\alpha$.   Thus the non-locality in a general theory can be determined  only by its local properties, namely, the balance relation.

As the first application we now examine the PR-box model within our framework. The PR-box has been actively  studied as a referenced  model in the exploration of physical principle specifying  QM \cite{ic,wvd}.   In this model the correlations assume the following form
\begin{equation}
p(a,b|A_\mu,B_\nu)=\frac{1+(-1)^{a+b+\mu\nu}}4,
\end{equation}
with $(a,b,\mu,\nu=0,1).$  For each measurement $B_\nu$ and outcome $b$ on Bob's side, a conditional state is prepared on Alice's side. And for each of the four states  prepared by Bob on Alice's hand, we always have  $\delta_{A_\mu}=0$ with $\mu=0,1$, i.e., the states  display no uncertainty while there is non-trivial complementarity in sequential measurements schemes \cite{gs}
  \begin{eqnarray}
 \delta_{A_0}+\delta_{A_1}=0, \quad  D_{A_0\rightarrow A_1}+D_{A_1\rightarrow A_0} >0. \label{prb0}
\end{eqnarray}
(See SM)
 Therefore, the PR-box has a vanishing balance strength $\alpha_{\rm pr}=0$ and it follows immediately from the upper bound Eq.(\ref{fia}) that PR-box has a largest possible nonlocal bound of $4$.
We observe that the PR-box  violates the balance relation
Eq.(\ref{qmr}), and the violation  reveals the discrepancy between local properties of QM and that of the PR-box.

As the second application we would like to derive the Tsirlsen bound. For this purpose we assume furthermore that the maximal violation to CHSH is attained when
$$\max\textstyle\left|\sum_{\mu,a,b}(-1)^{a+b+\mu\nu}p(a,b|A_\mu,B_\nu)\right|$$
is independent of $\nu$. From this additional symmetry assumption it follows that $f(\alpha,\gamma,\gamma)=f(\alpha,-\gamma,-\gamma)$ which gives $\gamma=0$ so that  the maximal violation Eq.(\ref{fia}) becomes
\begin{eqnarray}\label{p}
{\rm CHSH}_{s}\leq \frac{4}{\sqrt{\alpha^{2}+1}}\, ,
\end{eqnarray}
which is  shown  by the solid line labeled by $n_{s}$ in  Fig.1.
\begin{figure}
\begin{center}
\includegraphics[width=0.45\textwidth]{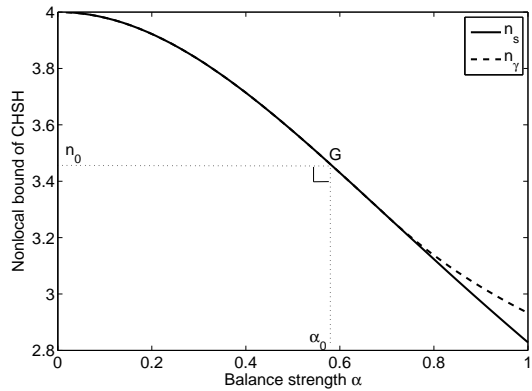}
\end{center}
\vspace{+0cm}
\caption{Relations between nonlocality upper bound and the balance strength $\alpha$. i) Balance relation
imposes a constraint on non-locality, and the non-local bounds are given  as functions of balance strength:    the dash line  $n_{\gamma}$ ($\alpha \leq 1$) represents the bound obtained under unbias assumption;  the solid line $n_{s}$ represents  the bound obtained   with the unbias assumption and the symmetry assumption; ii) Non-locality   also imposes a constraint on  balance strength: the point $G$ shows that   the balance strength of a theory which has  exhibited correlation strength $n_{0}$ should be no greater than $\alpha_{0}$. }\label{fig2}
\end{figure}
Taking into account the fact  that QM has a unit balance strength, i.e., $\alpha_{\rm qm}=1$, one can readily reproduce  Tsirelson bound from the upper bound Eq.(\ref{p}).
We note that without the additional symmetry assumption
 any theory  with unite balance strength  would have a nonlocality upper-bound of $n_{\gamma,\tau}=2.93$.

On the other hand, if the non-local bound of one theory is found to be $n_{0}$ as shown by the point  $G$ in  Fig.1, i.e., the maximal violation to CHSH inequality is $n_0$, then the corresponding balance strength of the given theory should be no greater than $\alpha_{0}$.     If  $\gamma=0$ assumption is taken again and from Eq.(\ref{p}) we obtain the following constraint of balance strength by nonlocality
\begin{eqnarray} \label{ss}
\alpha_{s}\leq \frac{\sqrt{16-n^{2}_{0}}}{n_{0}} .
\end{eqnarray}

\emph{Conclusion ---} We have introduced an uncertainty-complementarity balance relation, based on which we build   connections among non-locality, uncertainty, and complementarity.  Our considerations proceed without referring to any specific physical theory except the unbias  assumption. Therefore our results hold generally and can be used to specify nonlocal theories. As applications, we explain QM non-locality and the PR-box non-locality with  their  balance relations, respectively.

\emph{Acknowledgement ---}
This work has been supported by the Chinese Academy of Sciences, the National Natural Science Foundation of China under Grant No. 61125502, and
the National Fundamental Research Program under Grant No. 2011CB921300.

\newpage{\mbox{}}\newpage
\section{Supplementary Material}

{\it Proof of balance relation Eq.(\ref{qmr}) for QM--- } Suppose that a quantum mechanical system is prepared in the state $\rho$ and after an ideal Von Neumann measurement $\{\hat P_{j|A}=|a_j\rangle\langle a_j|\}$ is performed the system is brought into a completely decohered state $\rho_A$ in the $A$ basis. Denoting by $\{\hat P_{i|A^\prime}\}$ the measurement of $A^\prime$, we have
\begin{eqnarray}\nonumber
D_{A\rightarrow A'}
&=&\sum _{i}|\operatorname{Tr}(\rho-\rho_{A})\hat{P}_{i|A'}|\\ \nonumber
&\leq&\operatorname{Tr}|\rho-\rho_{A}|=\operatorname{Tr}\left|\textstyle\sum_{i< j}\sigma_{ij}\right|\\ \nonumber &\leq&\sum_{i< j} \operatorname{Tr}|\sigma_{ij}| =\sum_{i< j} 2 |\langle a_{i}|\rho|a_{j}\rangle|\\\label{3}
&\leq& \sum_{i< j} 2 \sqrt{p_{i|A} p_{j|A}}=\delta_{A}
\end{eqnarray}
where   $\operatorname{Tr}|X|:=\operatorname{Tr}\sqrt{X^\dagger X}$ and
 $\sigma_{ij}:=|a_{i}\rangle\langle a_{i}|\rho|a_{j}\rangle\langle a_{j}| + |a_{j}\rangle\langle a_{j}|\rho |a_{i}\rangle\langle a_{i}|$  for $i\not=j$. In the above proof, we have used the convexity of trace-norm in the second line.

{\it Proof the Balance strength  $\alpha_{pr}=0$ for PR-box  --- } Consider PR-box shared by Alice and Bob, by each measurement $B_{\mu}$  and outcome $b$ on Bob's side he would prepare a conditional state $\omega_{b|B_{\mu}}$ on Alice's side. By the non-signaling principle, Alice cannot  confirm which conditional state is prepared in her hand by local operation. Consider one case that Alice has obtained an outcome $a=0$ by measuring $A_{0}$ (output state would then be brought into  $\mathcal{S}_{0|A_{0}}$), she would know the measured state is either  $\omega_{0|B_{0}}$ or $\omega_{0|B_{1}}$.

 By $p(i|A_{1},\mathcal{S}_{0|A_{0}})$ and  $p(i|A_{1},\omega_{0|B_{0}})$  we denote the probabilities  of obtaining $i$ when measuring $A_{1}$ on $\mathcal{S}_{0|A_{0}}$ and on $\omega_{0|B_{0}}$.  Following the definition of  PR-box $p(1|A_{1},\omega_{0|B_{0}})=0, p(0|A_{1},\omega_{0|B_{0}})=1$, then disturbance $D_{A_{0}\rightarrow A_{1}}$ for $\omega_{0|B_{0}}$ is
\begin{eqnarray}\nonumber
D_{A_{0}\rightarrow A_{1}}(\omega_{0|B_{0}})
&=&\sum _{i}|p({i|A_{1}}, \omega_{0|B_{0}})-p(i|A_{1},\mathcal{S}_{0|A_{0}})|\\ \nonumber
&=&2(1-p(0|A_{1},\mathcal{S}_{0|A_{0}})).
\end{eqnarray}
Similarly, $D_{A\rightarrow A_{1}}(\omega_{0|B_{1}})=2p(0|A_{1},\mathcal{S}_{0|A_{0}})$, then
\begin{eqnarray}\nonumber
\sum_{i}D_{A_{0}\rightarrow A_{1}}(\omega_{0|B_{i}})=2.
\end{eqnarray}
By the definition of the box we have
\begin{eqnarray}\nonumber
\sum_{i}\delta A_{0}(\omega_{0|B_{i}})=0.
\end{eqnarray}
Following the general balance relation we have
\begin{eqnarray}\nonumber
\sum_{i}\delta A_{0}(\omega_{0|B_{i}}) \geq \alpha_{pr} \sum_{i}D_{A_{0}\rightarrow A_{1}}(\omega_{0|B_{i}}).
\end{eqnarray}
 Thus, $\alpha_{pr}=0$.

{\it Proof of nonlocality upper bound Eq.(\ref{fia})--- }
From the normalization conditions such as $\sum_iP({i}|A_1,\mathcal{S}_{\mu|A_0})=1$ and the unbias assumption Eq.(\ref{1}) for two sequential measurements $A_0$ and $A_1$
there is only two independent parameters, e.g., $\gamma$ and $\tau$ as given in Eq.(\ref{gt}),  among 4 probability distributions $\{P({i}|A_\nu,\mathcal{S}_{\mu|A_{1-\nu}})\}$ with $\mu,\nu=0,1$. The disturbance reads
\begin{eqnarray*}
 D_{A_{\mu}\to A_{\bar\mu}}&=& \sum _{a=0}^1\big|p({a|A_{\bar\mu}})-p({a|A_{\mu}\to A_{\bar\mu}})\big|\\
&=& 2\big|p({1|A_{\bar\mu}})-p({1|A_\mu\to A_{\bar\mu}})\big|\\
&=& 2\big|p({1|A_{\bar\mu}})-\textstyle\sum_a p({a|A_\mu})P(1|A_{\bar\mu},\mathcal S_{a|A_\mu})\big|\\
&=& \big|2p({1|A_\mu})-1+\gamma_\mu \overline A_{\bar\mu}\big|\\
&=& \big|\overline A_{\mu}+\gamma_\mu \overline A_{\bar\mu}\big|
\end{eqnarray*}
with $\mu=0,1$, $\bar\mu=1-\mu$, and
$$\gamma_\mu=P(1|A_\mu,\mathcal S_{0|A_{\bar\mu}})-P(1|A_\mu,\mathcal S_{1|A_{\bar\mu}}).$$
The first equality is the definition of disturbance; the second equality is due to the normalization of probability distributions; the third equality is the definition of $P_{a|A_\mu\to A_{\bar \mu}}$ as in Eq.(\ref{dis}); the fourth equality follows from the unbias assumption with $\gamma_0=\gamma$ and $\gamma_1=\tau$ as defined in Eq.(\ref{gt}); in the last equality we have used $\bar A_\mu=p({0|A_\mu})-p({1|A_\mu})$. As a result we have
\begin{eqnarray}\label{af}
\sqrt{1-\overline{A}_\mu^{2}}=\delta_{A_\mu}\geq \alpha D_{A_{\mu}\to A_{\bar\mu}}=\alpha\big|\overline A_{\mu}+\gamma_\mu \overline A_{\bar\mu}\big|
\end{eqnarray}
with $\mu=0,1$. Squaring Eq.(\ref{af}) and denoting $a=\overline{A}_0+\overline{A}_1$, $b=\overline{A}_0-\overline{A}_1$, we have
  \begin{align}\nonumber
   4\geq &a^{2}(\alpha^{2}(1+\gamma)^{2}+1)+
   b^{2}(\alpha^{2}(1-\gamma)^{2}+1)\label{n1} \\  &+2ab(\alpha^{2}(\gamma^{2}-1)+1)\\ \nonumber
   4\geq &a^{2}(\alpha^{2}(1+\tau)^{2}+1)+b^{2}(\alpha^{2}(1-\tau)^{2}+1)\\
      &-2ab(\alpha^{2}(\tau^{2}-1)+1)\label{n2}
\end{align}
from which we obtain, by eliminating $ab$ terms,
     \begin{eqnarray}
     4 \geq \frac{a^{2}}{f(\alpha,\gamma, \tau)}+\frac{b^{2}}{f(\alpha,-\gamma, -\tau)}\label{n3}
\end{eqnarray}
with $f(\alpha,\gamma,\tau)$ given by Eq.(\ref{f}).  As a result  it holds
 \begin{eqnarray}\label{aaa}
&&|\overline{A}_0\pm \overline{A}_1|\leq 2\sqrt {f(\alpha,\pm\gamma, \pm\tau)}
\end{eqnarray}
for any state of the subsystem in Alice's hand. We note that after Bob's measurement $B_\nu$ with outcome $b$ Alice's  subsystem will be brought into some state $\mathcal S_{b|\nu}$ with probability $p({b|B_\nu})$ so that $$p(a,b|A_\mu,B_\nu)=p({b|B_\nu})p(a|A_\mu,\mathcal S_{b|\nu}).$$ Denoting $\overline A_\mu(\mathcal S_{b|\nu})=\sum_a(-1)^ap(a|A_\mu,\mathcal S_{b|\nu})$ we have
 \begin{align}\nonumber
\operatorname{CHSH} &=\sum_{a,b,\mu,\nu=0}^1(-1)^{a+b+\mu\nu}p(a,b|A_\mu, B_\nu)\\ \nonumber
&\leq\sum_{ b,\nu=0}^1 p(b|B_\nu)\left|\overline{A}_0(\mathcal S_{b|\nu})+(-1)^\nu\overline{A}_1(\mathcal S_{b|\nu})\right|\\ \nonumber
&\leq  \sum_{ b,\nu=0}^1 p(b|B_\nu)2\sqrt {f(\alpha,(-1)^\nu\gamma, (-1)^\nu\tau)}\\
&=2\sqrt {f(\alpha,\gamma, \tau)}+2\sqrt {f(\alpha,-\gamma, -\tau)}
\end{align}
where the second inequality is due to Eq.(\ref{aaa}).
\begin{figure}[h]
\includegraphics[scale=0.33]{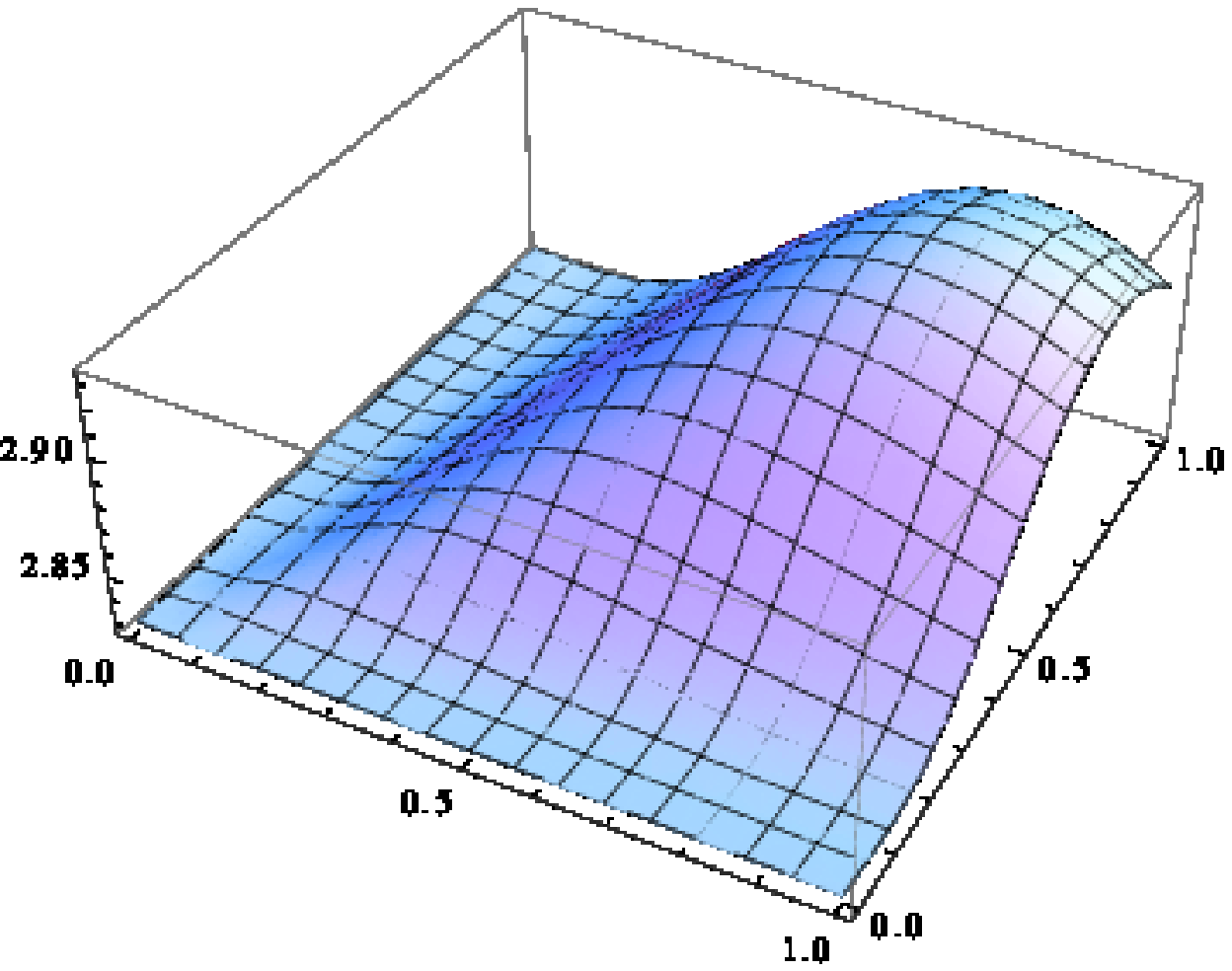}\includegraphics[scale=0.33]{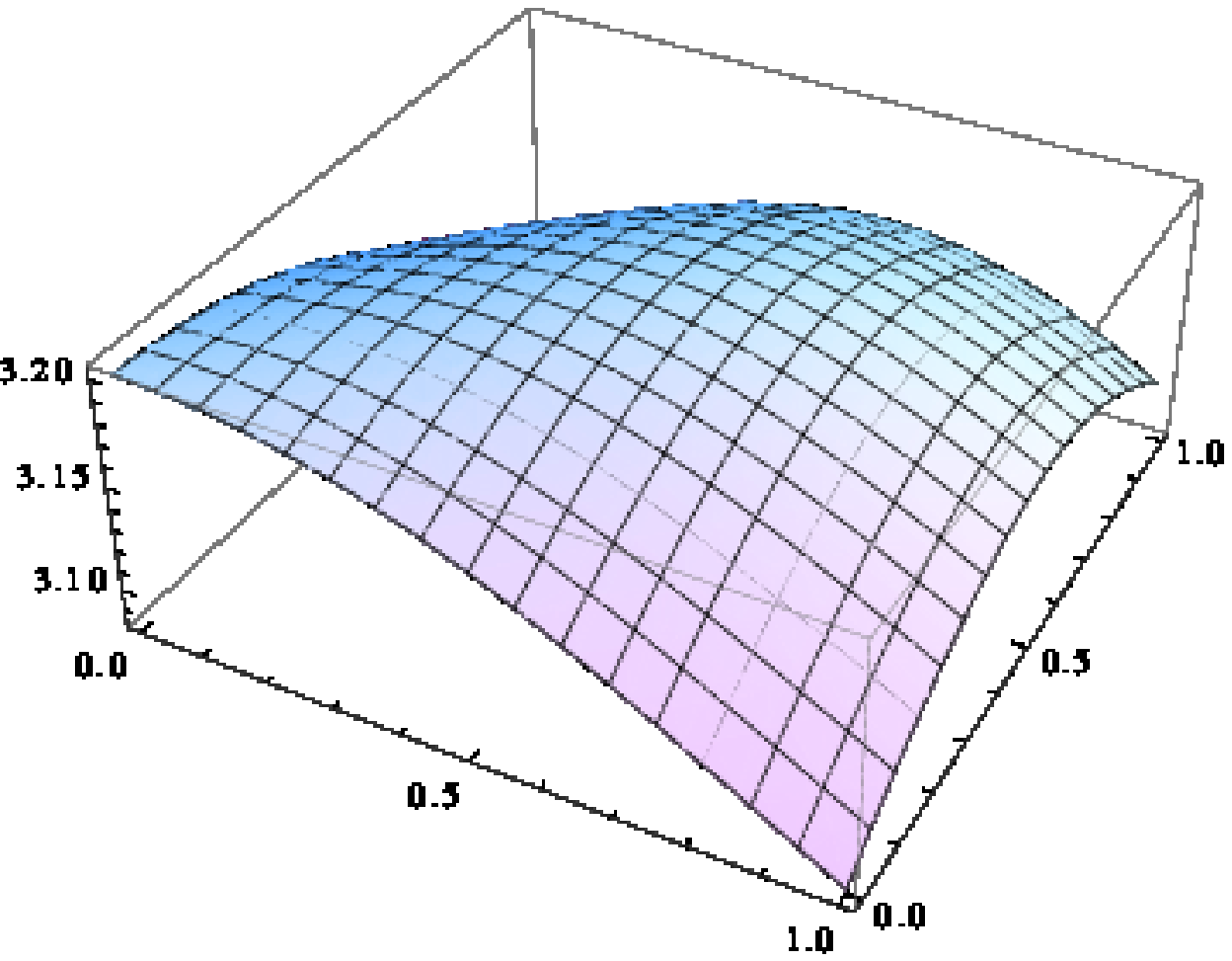}
\includegraphics[scale=0.33]{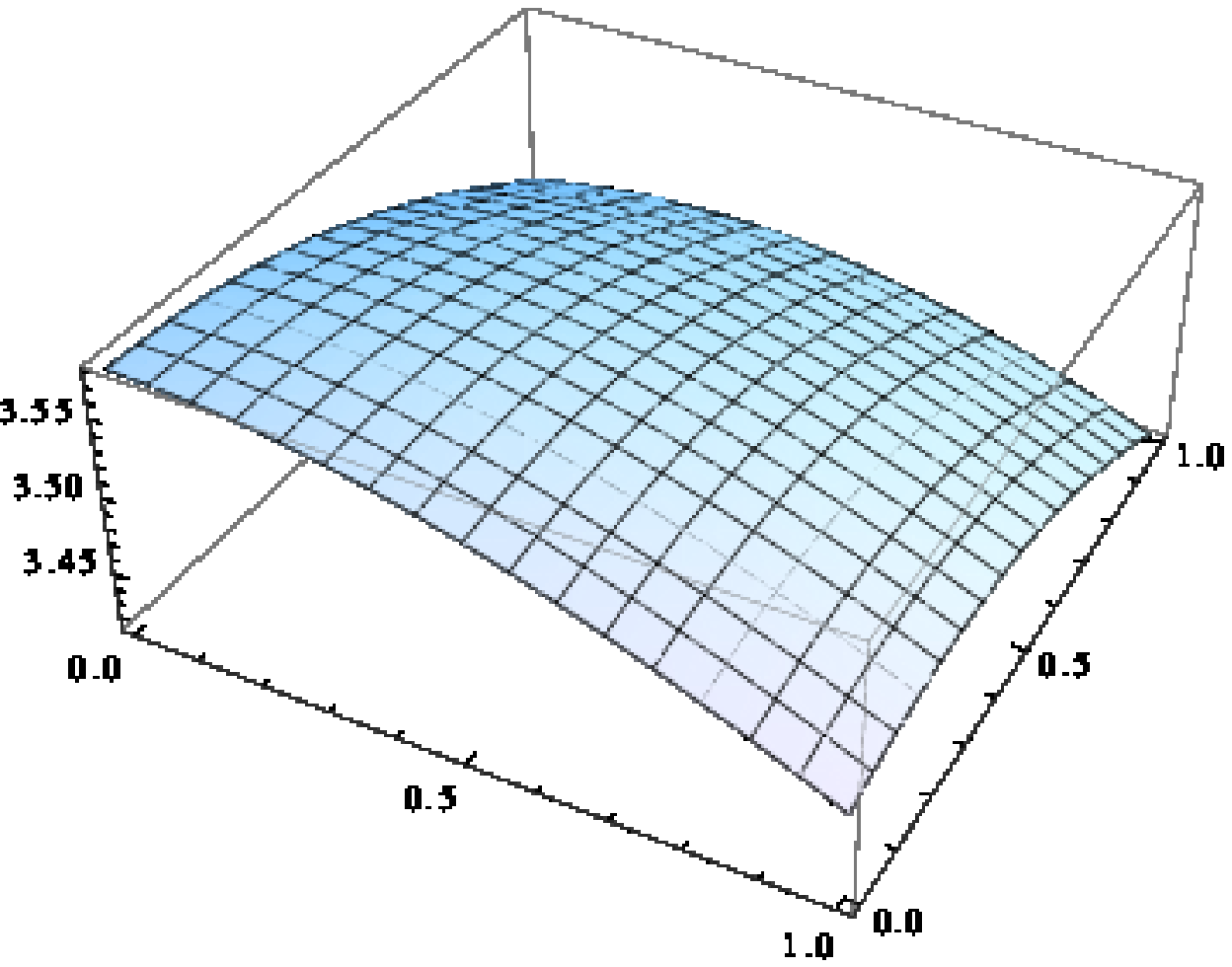}\includegraphics[scale=0.33]{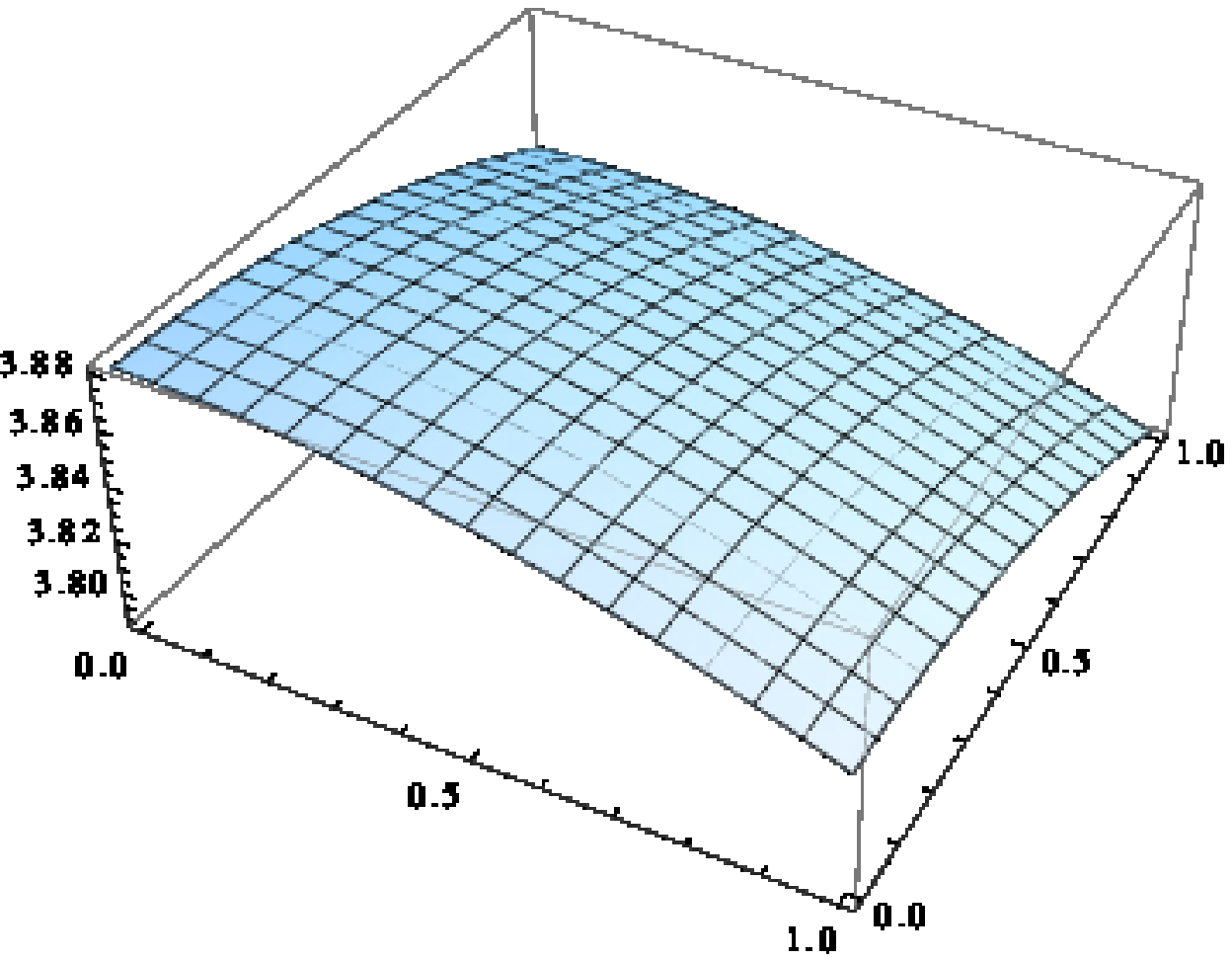}
\caption{(Color online) The nonlocality upper bound $n_{\gamma\tau}$ is plotted as the
 function of $\gamma$ and $\tau$ with $\alpha=1$ (upper left), $\alpha=3/4$ (upper right), $\alpha=1/2$ (lower left), and
  $\alpha=1/4$ (lower right). We see clearly that the maximum is always attained at $\gamma=\tau$.}
\end{figure}

\end{document}